# Hydrodynamic characteristics of sea kayak traditional paddles

**Pascal Hémon**

Hydrodynamics Laboratory (LadHyX), CNRS-Ecole Polytechnique, Palaiseau, France

Email: pascal.hemon@ladhyx.polytechnique.fr

Office phone: +33 1 69 33 52 76

**Abstract**

We present a study of the hydrodynamic characteristics of sea kayak paddles without taking into account the kayaker. We focus on traditional paddles used in the Arctic, one from Greenland and one from the Aleutian islands. A basic modern European paddle is included in the study for comparison. First the paddle stroke parameters specific to sea kayaking are identified because previous studies were devoted to a competition context. The hydrodynamic force generated by the blade motion is detailed: two terms are identified, one involving the inertia of the water surrounding the blade at the beginning of its motion, and the second term is the classical drag/lift force. Drag and lift force coefficients were measured in a wind tunnel. The data allow computation of the hydrodynamic force during a paddle stroke. The European paddle was shown to be more efficient than the traditional paddles because of its shorter length to width ratio which contributed to a larger inertia effect. However, the force obtained with the traditional paddles better follows the imposed motion by the kayaker so that they are more comfortable and less tiring in the context of long distance trips, as those practiced in sea kayaking.

**Acknowledgements**

The author is grateful to Caroline Frot from LadHyX for the 3D printing of the wind tunnel models and to Dr. Xavier Amandolese from LadHyX for the wind tunnel access and the force measurements. Traditional paddles have been manufactured and furnished by Alain Kerbiriou (www.kerlo.fr).

Preprint published in Sports Engineering, (2018) 21:189–197. doi.10.1007/s12283-017-0262-x## 1. Introduction

Hydrodynamics of kayak paddles have rarely been studied in the past and most studies are focused on biomechanics, taking into account the athlete physiology. In this paper we address the characteristics of paddles independent from the kayaker.

We focus on traditional paddles used in the Arctic. The kayak, and all the tools associated with it, have been developed and improved along centuries by Inuit and Greenlandic people [1]. They have acquired a strong traditional knowledge that has not been scientifically explained nor quantified. Kayaking has been exported outside Arctic as a sport, even to the Olympic games, or as a leisure activity for a larger public.

In the latter context, sea kayaking covers some practices that may be quite different, from the short journey during half a day, to long expeditions of several weeks. Some expert practitioners are interested in improving their techniques and their safety for sea kayaking. There is an interest in traditional paddles that are considered easier to use, safer and more comfortable during very long distance trips, than European paddles which design is generally inspired by competition practice [2, 3].

There are approximately two ways of using a paddle: the simplest and more common one is to maintain the blade normal to the flow along the stroke, which is performed by using a "drag paddle" because the generated hydrodynamic force is mainly a drag force. The other way of using a paddle is to apply an angle of attack to the blade, constant or not during the stroke, which makes the blade like a wing by creating a lift force normal to the blade direction of motion. To be propulsive, the kayaker must adapt his paddle stroke and is generally limited to expert or elite kayakers [4].

There is a need for improving the knowledge about kayak paddles, especially in the context of sea kayaking, while the competition context has been investigated [5]. Here we focus on the hydrodynamic force that is obtained on the blade paddle during the stroke. The influence of the design parameters of sea kayak paddles has not been clarified and a careful hydrodynamic investigation is needed.

The paper is organized as follows. First, we describe the paddles in § 2. Then the paddle stroke parameters of a typical sea kayaker are measured in § 3. The hydrodynamic force is detailed and experiments in a wind tunnel are performed in § 4. Paddle performance is compared in § 5 against a simplified paddle stroke cycle.

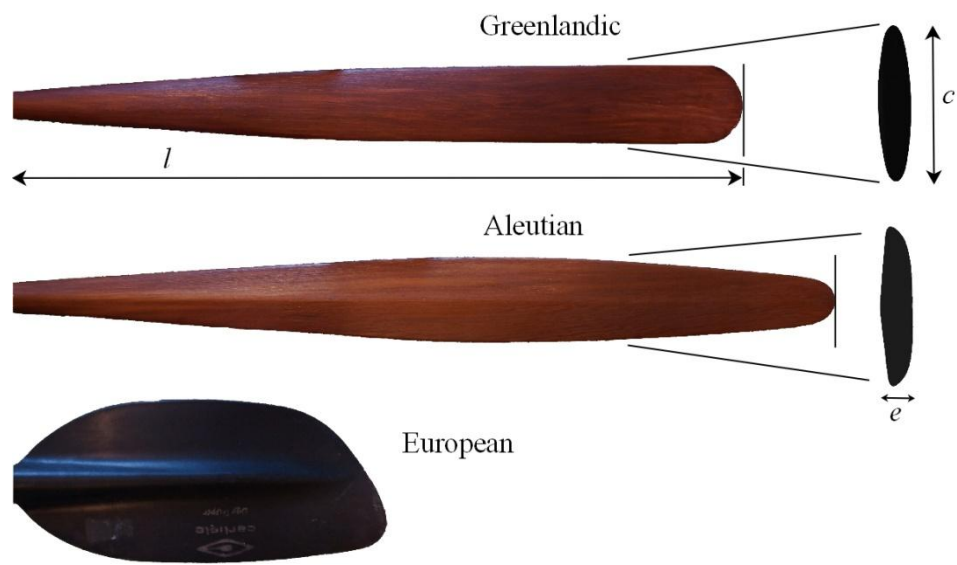

**Fig. 1** View of the paddle blades with relative scale respected. Transversal shape at mid-length is shown for the Greenlandic and the Aleutian blades

## 2. The Greenlandic and Aleutian paddles

There are numerous types of traditional paddles because people of the Arctic have developed their tools in deep relation with their environment so that each community has its own design adapted to the local sea conditions. A review of paddles has been realized in [6, 7, 8].

It is not possible to study all the paddle types. Two traditional designs are chosen: a Greenlandic paddle, with symmetric blades and an Aleutian paddle for which the blades have an extrados different from the intrados. In particular the extrados has a characteristic flat dihedral shape. A third basic "European" paddle is included in the study for comparison. These paddles are presented in Fig. 1.

The main characteristics of the paddles are given in Table 1 where blade dimensions used further are detailed: $e$ is the mean thickness, $c$ the maximum width or chord, $c_h$ the chord at 1/3 from the tip, $l$ the blade length and $S$ the blade surface (projected area). The latter was measured with calibrated pictures (ImageJ). Differences in dimensions appear between the Greenlandic and the Aleutian paddles due to local sea conditions which are rougher around Aleutian islands. The Aleutian paddle is designed to obtain a larger force through a longer shaft and a larger blade surface.

The two traditional paddles are manufactured from western red cedar (Thuja Plicata). Specific gravity is 0.37-0.38 and Young modulus is in the range 7900-8800 MPa. The external surface is oiled for better protection against water. The European blade is made of plastic and the shaft of aluminum.

**Table 1** Characteristics and dimensions of the studied paddles

| Paddle | Mass (kg) | Length (m) | Shaft diameter (m) | $e$ (m) | $c$ (m) | $c_h$ (m) | $l$ (m) | $S$ (m$^2$) | $l/c_h$ |
|---|---|---|---|---|---|---|---|---|---|
| Greenlandic | 0.86 | 2.06 | 0.035/0.029 | 0.007 | 0.088 | 0.088 | 0.78 | 0.0560 | 8.9 |
| Aleutian | 0.94 | 2.30 | 0.0375/0.030 | 0.012 | 0.095 | 0.085 | 0.85 | 0.0644 | 10 |
| European | 0.98 | 2.20 | 0.030 | 0.004 | 0.181 | 0.181 | 0.41 | 0.0597 | 2.3 |

## 3. The paddle stroke parameters

We assume in this paper that the kayak velocity has reached a stationary value $U$, constant, which corresponds to a cruise regime see Fig. 2. There is then a cycle of the paddle velocity $v$ in water that has to be clarified. The evolution of $v$ during the stroke can be assimilated roughly by a half-period of a sinusoid, [9]. Only two parameters are sufficient to characterize this stroke cycle: the duration of the stroke $T$ when the blade is wetted and the maximum velocity $v_{max}$ reached by the blade.

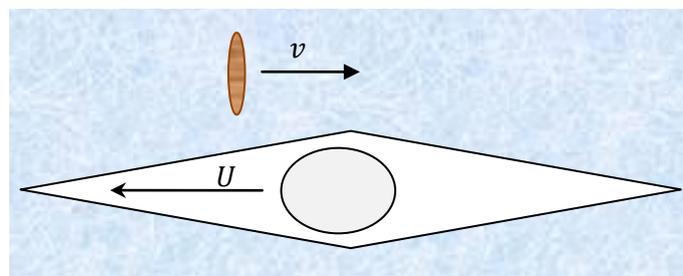

**Fig. 2** Definition of the kayak velocity $U$ and the paddle velocity $v$

In the paper of Caplan [9], estimation of $T$ gives 0.4 s and $v_{max} = 1.23$ m/s. Others have found $0.3 < T < 0.59$ s [10-12]. In [13] $v_{max} = 3$ m/s. These values concern elite kayakers using European paddles so that their extrapolation to the traditional sea kayak practice seems inappropriate.

**3.1 Duration of a paddle stroke**

Measurements are made using embedded cameras: one fixed on the kayak roof and another fixed on the shaft at the bottom of the paddle blade, as shown in Fig. 3 (60 fps). The cameras were synchronized by a "clap" at the beginning of the record. Tests were performed on sea, in the Morbihan Gulf in France that offers very flat water conditions.

Ten consecutive paddle strokes are analyzed in the video sequence and during cruise conditions where the hull velocity is constant and in a straight line. The mean duration of the stroke is found to be $T = 0.73 \pm 0.03$ s which is longer than that found by others using European paddle [10-13].

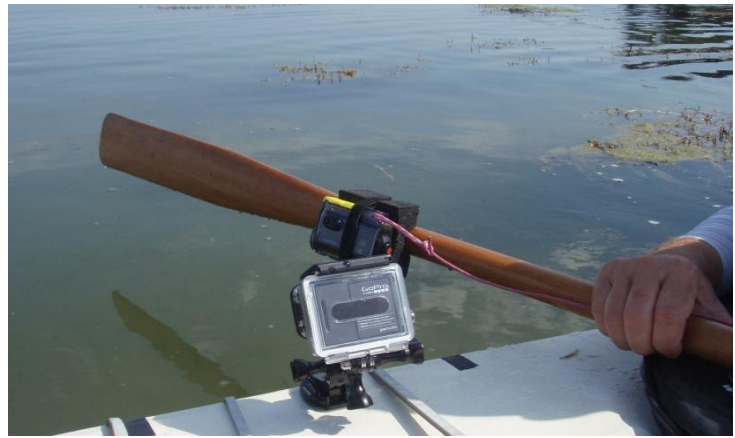

**Fig. 3** A view of the cameras on the kayak roof and fixed on the paddle

**3.2 Velocity of the paddle blade in water**

The velocity of the paddle blade in water were performed using a flow meter mounted on the paddle as shown in Fig. 4. The lateral distance between the flow meter and the blade is twice the width of the blade to avoid interaction between them. Alignment of the flow meter with the blade was performed on a flat table. They are linked together with two clamps screwed on the paddle. The sensor is a FLOW PROBE type FP111 which is composed of a small rotor and a digital recorder linked by a shaft. Accuracy is 0.03 m/s in the range 0.1-6.1 m/s. The directional sensitivity was verified and found to be negligible inside a cone of 12°. The recorder can display the maximum value seen by the sensor which was measured over ten paddle strokes.

Results of measurements gave: $v_{max} = 0.70 \pm 0.1$ m/s which is less than reported elsewhere [9, 13]. The velocity of the hull was found by GPS to be around 4 knots (2 m/s) which corresponds to a common cruise speed.

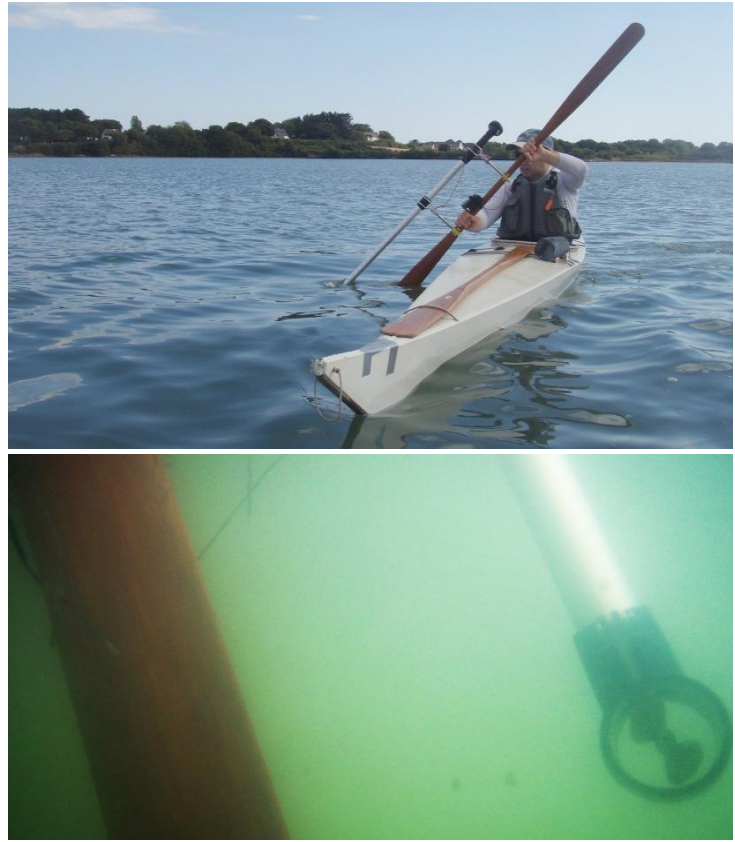

**Fig. 4** Views of the flow meter mounted on the paddle

## 4. Hydrodynamic forces

Jackson et al. [5] presented a hydrodynamic study of modern European paddles by comparing the classical "drag paddle" with the "wing paddle". Here we focus on the drag paddle. During the cruise regime, the drag paddle keeps its blade normal to the flow and the motion is parallel to the hull axis, see Fig. 2. The hydrodynamic force applied to the blade, which becomes the propulsive force through the kayaker [14], can be written as

$$|F_x| = m_a\, \sigma(T_e)\, \gamma + \frac{1}{2} \rho\, S\, C_d\, v^2 \qquad (1)$$

where, $\gamma$ and $v$ are the acceleration and the velocity of the blade in water, $m_a$ the added mass, $\sigma(T_e)$ a weight function depending on the establishment time $T_e$, $C_d$ the blade drag coefficient and $\rho$ the water density.

The hydrodynamic force is then decomposed into two terms detailed hereafter. The term $m_a\, \sigma(T_e)\, \gamma$ is due to the inertia of the water around the blade that has to be "pushed" by the blade, and the term $\frac{1}{2} \rho\, S\, C_d\, v^2$ is the classical drag/lift force.

### 4.1. Inertial term

There are different approaches to take into account the inertia of the water surrounding the blade. Jackson [5] used the concept of starting vortices to develop an expression of this force. This expression is however not completely applicable here because the traditional paddle blades are much longer than large, whereas this is not the case of the European paddles for which the expression was developed.

The other way to identify the effect of the water inertia is the concept of added mass which is widely used in fluid-structure interactions and vibrations studies in offshore industry [15]. Added mass can be seen as the mass of the surrounding water which is put into motion by the blade. Blevins [15] proposed an analytical expression of the added mass for flat plates perpendicular to a flow and having different length to width ratios. It is

$$m_a = \frac{\pi}{4} \rho \, l \, c_h^2 \qquad (2)$$

In the present case, the blades are not real rectangular plates so that the above expression is modified in

$$m_a = \frac{\pi}{4} \rho \, S \, c_h \qquad (3)$$

which can take into account the real surface of the blade in place of $l \, c_h$. This added mass is given in the Table 2 for the three considered blades. We see the effect of a large chord that leads to a large added mass for the European paddle.

One of the assumptions yielding to (3) is a small displacement of the structure which is not the case during a paddle stroke cycle. From the parameters identified above the distance traveled by the blade is around 0.5 m which is almost 5-6 times the chord, whereas it is commonly admitted that the added mass expression cannot be used directly for a displacement greater than one chord.

A correction has therefore to be achieved to limit the effect of added mass during the beginning of the paddle stroke. Experiments published in [16, 17] give the hydrodynamic force on translated plates versus a non-dimensional time $T_e$ given by

$$T_e = \frac{1}{c} \int_0^t v(\tau) \, d\tau. \qquad (4)$$

This time $T_e$ was identified as a universal time scale for vortex ring formation [18]. In a sense, it corresponds to the distance in chord traveled by the paddle blade. Force measurements showed that for $T_e > 1.8 - 2$ the force has reached its constant value after the acceleration period. The effect of added mass then becomes negligible after this time. It is possible to construct a weight function $\sigma(T_e)$ applied to the added mass which is 1 at $T_e = 0$ and falls to zero at $T_e = 1.8$. In this paper a simplified approach is developed for the paddle stroke so that the simplest evolution, linear, was chosen for $\sigma(T_e)$.

**Table 2** Added mass of the studied paddle blades

| Paddle | $c_h$ (m) | $S$ (m$^2$) | $m_a$ (kg) |
|---|---|---|---|
| Greenlandic | 0.088 | 0.0560 | 3.87 |
| Aleutian | 0.085 | 0.0644 | 4.30 |
| European | 0.181 | 0.0597 | 8.49 |

**4.2. Blade profile characteristics**

To identify the second term of the hydrodynamic force, the drag coefficient $C_d$ of each blade has to be quantified. In [19] wind tunnel tests are presented concerning the drag and lift coefficients of rectangular flat plates of different length to chord ratios. A study in 1995 [20] presents results for common European paddles. More recently, European blades of different designs were also tested in wind tunnel [13]. A short report [21] concerning the Greenlandic blade was published. However, data are not complete and a systematic wind tunnel tests program is consequently presented further.

The blades have a transverse profile and a 3D shape which can be studied separately. Two series of tests were performed: one with 2D profiles of the traditional blades, see Fig. 2, and a second with 3D models of the blades. In each case, the angle of attack is the variable parameter of the tests, with the conventional definition shown in

Fig. 5. In the simplest paddle stroke using a drag paddle, the drag force at +90° is concerned. However, measurements are performed to provide the drag and lift force coefficients versus the angle of attack.

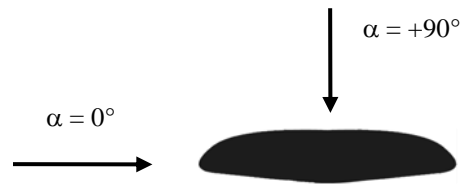

**Fig. 5** Definition of the angle of attack on the Aleutian blade paddle

The wind tunnel models are made via 3D printing based on coordinates measured on the real paddles. The scale of the models is 1/2 for the 2D profiles and 1/4 for the 3D blades. These models are shown in Fig. 6. The 2D profiles are tested between walls and have a length-to-chord ratio of 4 which is considered acceptable [22].

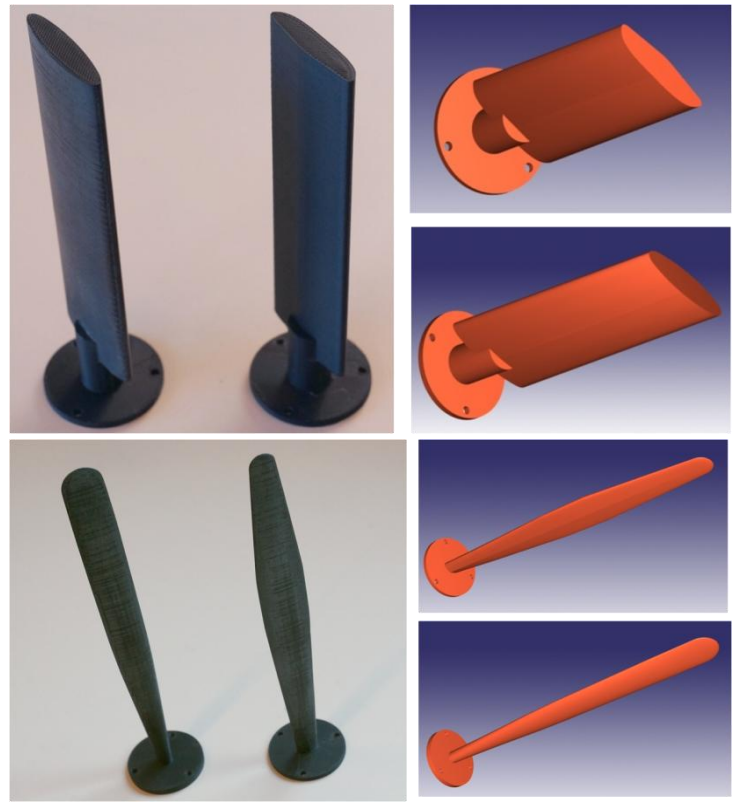

**Fig. 6** Views of the 2D (top) and 3D (bottom) wind tunnel models.
Left are photos of the printed models and right are the computer models.

Measurements are performed in a horizontal wind tunnel of the laboratory by means of a six components force sensor (Type NANO 43 from ATI Industrial Automation) and acquisition signals with a PAK system from Muller-BBM. The records are 10 s long sampled at 1024 Hz. Only time average values of forces are presented hereafter which accuracy is better than 5 %. The angle of attack is obtained via a motorized system which is controlled by a program defined in advance. Accuracy is of about 1°.

Other wind tunnel parameters are the reference velocity which is measured via a differential static pressure measurement between the settling chamber and the test section (KIMO type CP303). Speed was found using Bernoulli's equation, where temperature and atmospheric pressure where accounted for. Global accuracy of the velocity is under 1 %. The maximum solid blockage ratio $b$ of the test section is 4.8 % with the 3D models oriented at 90° and 6.5 % with the 2D models. The reference velocity is corrected by multiplying the raw value by the coefficient $(1 + b \sin \alpha)$, to take into account of the flow acceleration around the model. No other corrections are performed.

Reynolds number for the tests is $Re = \frac{v\, c_h}{\eta} = 40000$ where the velocity $v$ is 2/3 of the paddle velocity $v_{max}$ identified previously and $\eta$ is the kinematic viscosity (15 $10^{-6}$ m$^2$/s in air and $10^{-6}$ m$^2$/s in water [22]). In wind tunnel investigations, the Reynolds number is used for quantifying the effect of scale and a change in the fluid nature. Here the Reynolds number in the wind tunnel with air and scaled models is the same as the Reynolds number with real blades in water. This gives a wind tunnel velocity $v = 15$ m/s for the 2D profiles tests and 30 m/s for the 3D models.

Preliminary tests for another Reynolds number showed a small influence for the recorded drag force evolution and almost no influence for the lift force. Note that the blade models have a small surface roughness (< 40 μm) which ensures the turbulent character of the boundary layer as in a real paddle stroke.

The results are presented in Fig. 7 for the Greenlandic paddle and in Fig. 8 for the Aleutian paddle. $C_d$ is the drag coefficient and $C_l$ the lift coefficient. The reference surface in the case of the 3D blade is the projected area of the model measured by calibrated pictures.

There are not many differences between the 2D blade profiles: the drag at +90° is 1.74 for both. In the same conditions a flat plate drag coefficient was measured at 1.94 ± 0.09, which agrees with the literature [15,19] and validates the testing procedure. However for the 3D blades, the difference is noticeable, 1.43 for the Greenlandic and 1.59 for the Aleutian, that is 10 % larger. The 3D shape of the Aleutian blade, thinner at its tip is favorable. Note the non-symmetric curves for the Aleutian paddle due to its transversal shape, having a drag coefficient at -90° of 1.49. For comparison, the European blade in [13] has a drag coefficient of 1.70, slightly larger. These data are summarized in Table 3.

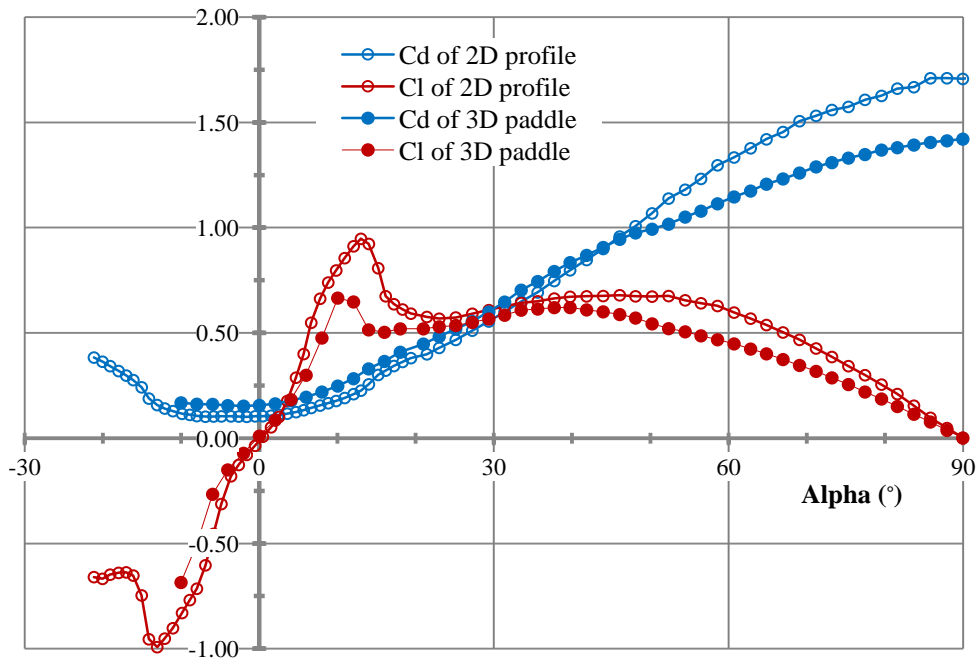

**Fig. 7** Drag and lift coefficients for the Greenlandic 2D profile and the 3D blade versus angle of attack

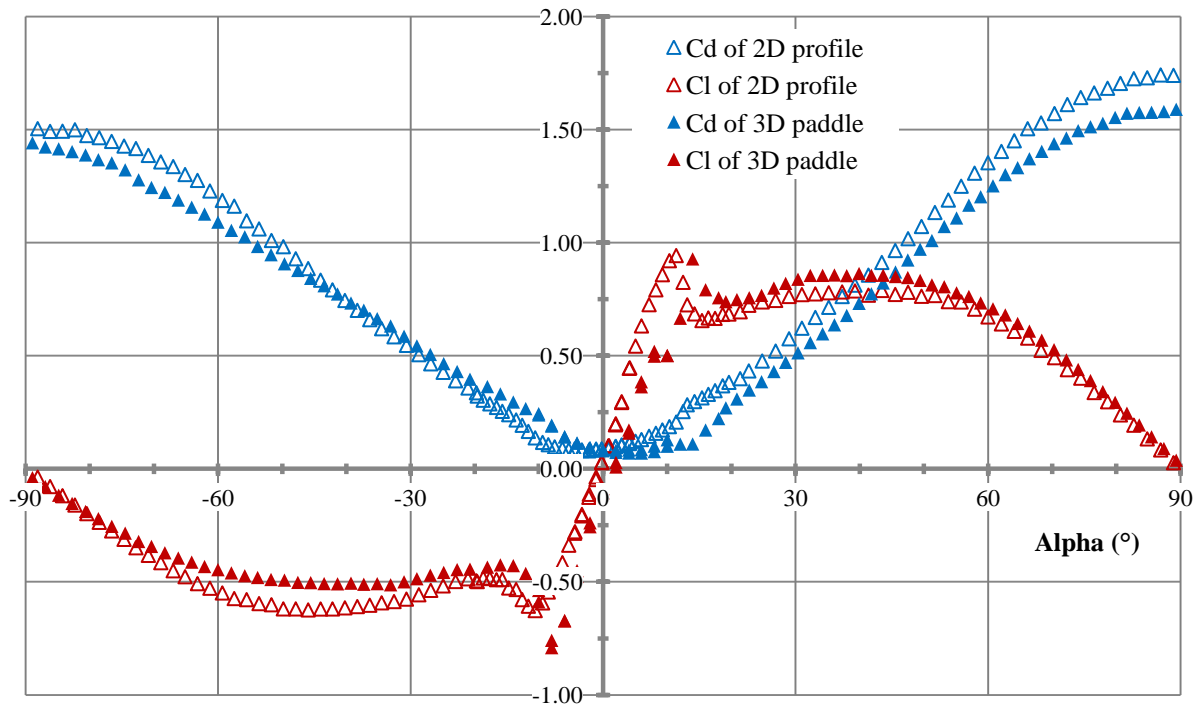

**Fig. 8** Drag and lift coefficients for the Aleutian 2D profile and the 3D blade versus angle of attack

**Table 3** Drag coefficients of the 2D profiles and 3D blades at +90°

| **Paddle** | 2D profile | 3D blade |
|---|---|---|
| Greenlandic | 1.74 ± 0.09 | 1.43 ± 0.07 |
| Aleutian | 1.74 ± 0.09 | 1.59 ± 0.08 |
| European [13] | - | 1.70 |

It appears that the differences in the drag coefficients come from the 3D shapes of the blades, but not from their transverse profiles which influence is only seen on the lift coefficients. Lift of a paddle can be important when the paddles are not used in a pure drag propulsion manner or for rolling maneuvers. For comparison, the lift coefficient of a flat plate and of a common wing section (NACA 0012) are shown in Figure 9 together with the lift coefficient of the traditional profiles. It appears that in the range 0-12° (mainly rolling maneuvers) the lift coefficients are almost similar. Note also the good behavior of the flat plate for which the lift coefficient does not present an abrupt stall around 15° as this is the case for the other profiles. Around +90° the flat plate is slightly better than the traditional profiles but remains close. Unfortunately, it is not possible to build a sufficiently stiff paddle with a thin flat plate as the one tested here.

The comparison of the 3D blades drag and lift coefficients is shown in Fig. 10. It appears that the European blade is close to the Aleutian blade. The Greenlandic blade has lower drag than the others but its symmetrical shape brings other advantages, especially during rolling maneuvers, when the kayaker has no choice in urgency for taking over the paddle.

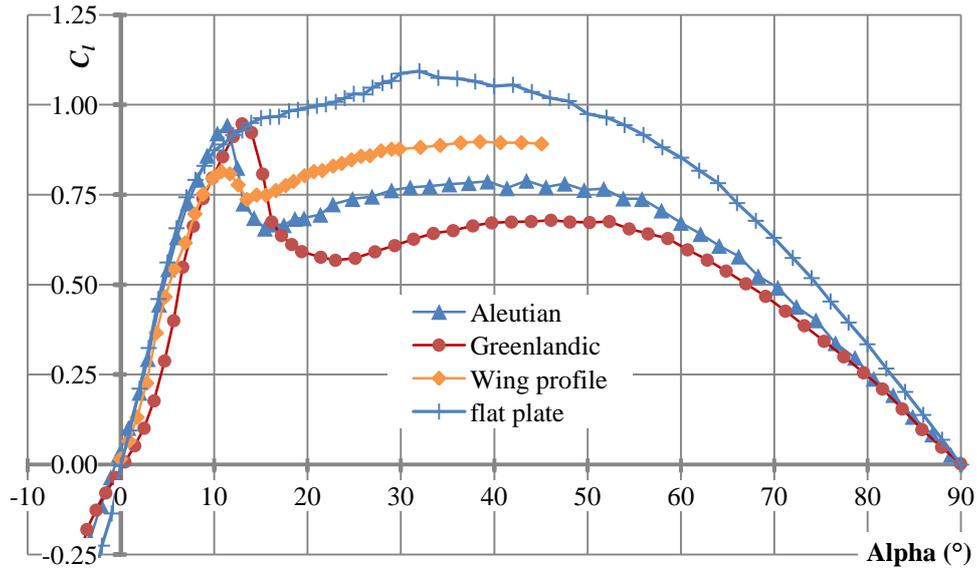

**Fig. 9** Comparison of the lift coefficient for different 2D profiles versus angle of attack

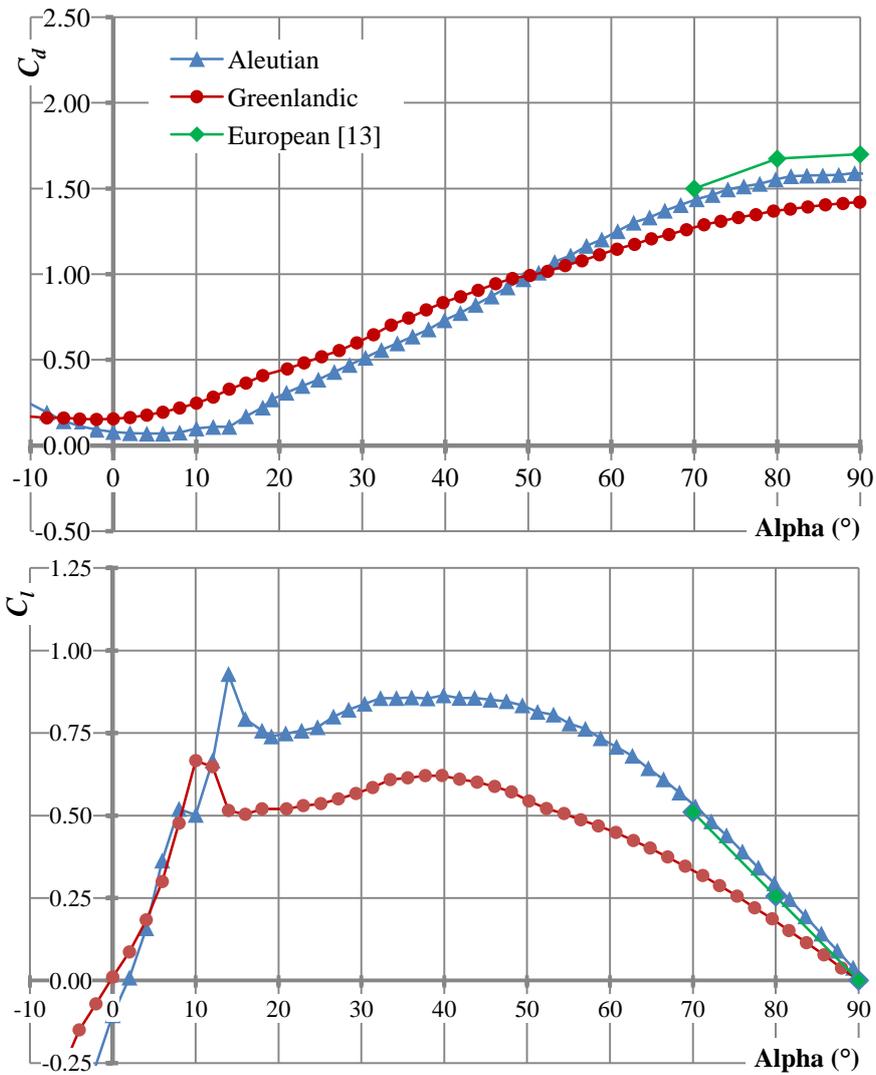

**Fig. 10** Comparison of drag (top) and lift (bottom) coefficients for the 3 blades versus angle of attack

## 5. Performance comparison

In this section we use the previously determined characteristics to compute the hydrodynamic force for each paddle. Measurements of force have been already published in [23] but only for an elite kayaker, i.e. in a competition context.

Here the paddle stroke is simplified to a sinusoid cycle during which the blade is supposed to start to move when it is entirely into water and stops in water. The phases of entering and exit of water are then not studied here. Moreover the blade is maintained normal to the flow (angle of attack +90°) and it is supposed to have only translational motion. This is a simplified paddle stroke which however allows comparisons for different paddles without any influence of the kayaker abilities in paddling with paddles of different length and type.

The blade motion follows an imposed evolution such that

$$v = v_{max} \sin\left(\pi \frac{t}{T}\right) \qquad (5)$$

and by consequence the acceleration is

$$\gamma = \frac{\pi}{T} v_{max} \cos\left(\pi \frac{t}{T}\right) \qquad (6)$$

where the dimensional time $t$ is taken from 0 to $T$. By using the different parameters of the stroke cycle found in § 3 and those of the hydrodynamic force in § 4 it is possible to compute the evolution of the force produced on the blade versus time. These results are shown in Fig. 11.

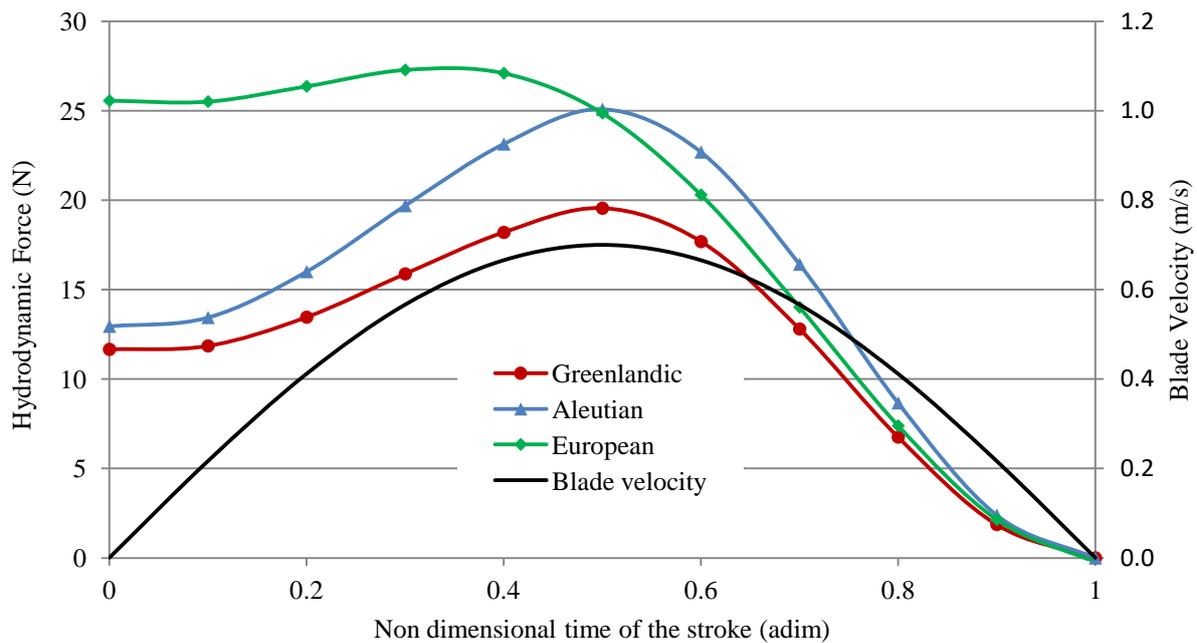

**Fig. 11** Comparison of the hydrodynamic force for the 3 paddles along a standard stroke

At the beginning of the stroke the force is dominated by the inertial term involving the added mass. Due to its large chord, the European paddle presents immediately a larger force. The influence of the chord is in fact double: obviously the added mass is larger, and secondly the duration $T_e$ is longer. For the traditional paddles the maximum force is obtained later during the stroke because the force is dominated by the drag term that follows the imposed velocity.

The Aleutian paddle provides a larger force than the Greenlandic paddle because the paddle surface is larger (see Table 1). It is however interesting to compare all these paddles independently of their blade surface. The non dimensional impulse which is presented in [24] as

$$I = \frac{\frac{1}{T}\int_0^T F(t)dt}{\frac{1}{2}\rho S v_{max}^2} \qquad (7)$$

provides a single numerical result which characterizes each blade and is given in Table 4. Uncertainties correspond to a maximum ±5 % of error on the drag coefficient. We see then that the European paddle is more efficient than the traditional paddles. Also we observe that the Aleutian is slightly better than the Greenlandic but not as much as it was expected with dimensional force which included the blade surface.

**Table 4** Non dimensional impulse of the 3 paddles

| Paddle | $I$ |
|---|---|
| Greenlandic | 12.9 ±0.5 |
| Aleutian | 13.9 ±0.5 |
| European | 18.8 ±0.5 |

## 6. Conclusion

A study of the hydrodynamic force on sea kayak traditional paddles has been presented. The paddle stroke parameters specific to this sport have been measured in a realistic environment. The hydrodynamic force on the blades is written, as the sum of an inertial term and of a drag term. The inertial term is modeled using the concept of added mass in which the blade width, the chord, has a major influence: the larger it is, larger is the resulting force. The drag/lift coefficients of the blades have been identified in a wind tunnel as a function of the angle of attack. Starting from a simplified stroke cycle, the force was calculated.

It appears that the European paddle is more efficient than traditional paddles. However, in the context of long trip with sea kayak, the traditional paddles are more comfortable because the force cycle better follows the motion cycle imposed by the kayaker. The added mass of traditional paddles is much smaller than for the European paddle so that the inertial force is weaker and shorter. The Aleutian paddle was found slightly more efficient than the Greenlandic paddle which however has the advantage of being fully symmetric and easier to take over.

Although only the drag is used in the standard paddle stroke, the lift force was also measured to provide data that could help to better model the paddle stroke which is used by an expert kayaker. Extension can also be made to sprint race on the basis of the paddle motions.

Improvements of the inertial term of the hydrodynamic force can be done by using a measured added mass, adapted to the 3D shape of the blades, instead of using the analytical (and simplified) expression which was used in this study.